# Reviewing local and integrated energy system models: insights into flexibility and robustness challenges

Febin Kachirayil[a, *], Jann Michael Weinand[b], Fabian Scheller[c], Russell McKenna[a,d]

[a] Department of Mechanical and Process Engineering, ETH Zürich, Switzerland

[b] Institute of Techno-economic Systems Analysis (IEK-3), Forschungszentrum Jülich GmbH, Germany

[c] Energy Economics and System Analysis, Division of Sustainability, Department of Technology, Management and Economics, Technical University of Denmark (DTU), Denmark

[d] Laboratory for Energy Systems Analysis, Paul Scherrer Institute, Switzerland

## Abstract

The electrification of heating, cooling, and transportation to reach decarbonization targets calls for a rapid expansion of renewable technologies. Due to their decentral and intermittent nature, these technologies require robust planning that considers non-technical constraints and flexibility options to be integrated effectively. Energy system models (ESMs) are frequently used to support decision-makers in this planning process. In this study, 116 case studies of local, integrated ESMs are systematically reviewed to identify best-practice approaches to model flexibility and to address non-technical constraints. Within the sample, storage systems and sector coupling are the most common types of flexibility. Sector coupling with the transportation sector, specifically with electric vehicles that could be used for smart charging or vehicle-to-grid operation, is rarely considered. Social aspects are generally either completely neglected or modeled exogenously. Lacking actor heterogeneity, which can lead to unstable results in optimization models, can be addressed through building-level information. A strong emphasis on cost is found and while emissions are also frequently reported, additional metrics such as imports or the share of renewable generation are nearly absent. To guide future modeling, the paper concludes with a roadmap highlighting flexibility and robustness options that either represent low-hanging fruit or have a large impact on results.

## Highlights

- Review of flexibility and robustness aspects of local, integrated energy models
- Flexibility comes from storage and sector coupling, demand-side management is rare
- Transport integration and the flexibility potential of EVs receive little attention
- Reporting metrics beyond cost and emissions provides additional decision-support
- Uncertainty assessments are overwhelmingly deterministic, mainly using scenarios

## 1 Introduction

There is wide consensus about the need to mitigate the effects of climate change by decarbonizing the energy system. Fossil-fuel dependent sectors such as transportation or heating require alternative energy carriers and a promising candidate for this shift is electrification [1]. This implies a massive expansion in renewable generation capacities and introduces a need for flexibility options due to the decentral and intermittent nature of renewable technologies such as solar photovoltaics (PV) or wind power as electricity demand has to be met instantaneously [2].

---

* Corresponding author. *E-Mail address:* fkachirayil@ethz.ch

Energy system models (ESMs) have been developed to support decision-makers in planning energy transitions. Since their inception, it has been argued that ESMs should be used for "insights, not numbers" [3]. To ensure robust insights, models must provide an accurate representation of the real-world system, including the techno-economic dimension, but also social, environmental or political constraints [4]. Uncertainties are another critical element to consider as decision-maker priorities shift over time and ESMs cannot reproduce all real-world processes and interactions [5].

A key trend of the energy transition is the emergence of an increasingly decentralized energy system as renewable energy sources exhibit lower energy densities. This shift can provide benefits to communities, from increased resilience and flexibility of their energy supply to direct involvement in the decision-making process [6]. Taking advantage of these opportunities requires planning on a local level and with local flexibility provision in mind [4,7].

The main challenges of local ESMs can largely be separated into two areas, as discussed insection 2. Firstly, the increasingly intermittent energy supply requires not only a high temporal and spatial resolution to identify the challenges resulting from this shift, but also an appropriate representation of technical flexibility to discover how to effectively overcome them [8]. Second, to ensure robust planning of an increasingly complex energy system, it is essential to tackle uncertainties, but also non-technical constraints such as heterogeneity, behavior or acceptance [9].

The aim of this paper is to describe the state-of-the-art modeling of different options for technical flexibility and robustness in local, integrated energy system models. To this end, a systematic literature research was performed, followed by a manual filtering of the resulting studies. In the latter step, studies were selected based on the use of a systemic view and the consideration of at least the power and thermal sectors to guarantee comprehensive and integrated energy system planning. Furthermore, a real-world application and investment planning were additional conditions to ensure that the sample provides insight to decision-making.

The final selection of 116 articles was analyzed using an attribute catalog that considers technical flexibility and robustness aspects in addition to more general information on the model structure. Technical flexibility was divided into five dimensions, supply- and demand-side options, storage, networks and sector coupling, and robustness into the assessment of uncertainty and non-technical constraints. With this approach, best practices are identified and unresolved challenges are highlighted. This study thereby represents the first systematic review of local, integrated ESMs focusing on the implementation of technical flexibility and robustness.

The rest of the article is structured as follows: section 2 provides a literature review, discussing the modeling of flexibility and robustness before delving into the challenges with local energy systems modeling. The review methodology is presented in section 3, separated into the article selection process and the study evaluation. Section 4 presents the results of the analysis, focusing on the implementation of technical flexibility and robustness options and section 5 discusses these results before section 6 provides a summary and a research roadmap based on the results of the review.

## 2  Literature Review

Sections 2.1 and 2.2 show why flexibility and robustness, respectively, are important components of ESMs and discuss prior work addressing these dimensions, before section 2.3 highlights our contribution relative to the previous work.

## 2.1 Flexibility modeling

Flexibility is an essential aspect of energy systems and especially of power systems, where demand has to be met instantaneously [2]. With the emergence of intermittent renewables, power supply is becoming more volatile, so that methods to address mismatches between generation and supply gained importance [10]. Options for flexibility include supply- and demand-side measures to shift the timing or amount of generation or consumption, storage to address temporal mismatches or distribution networks for spatial disparities [11]. In addition, sector coupling of the power sector with heating, cooling or transportation can be used to obtain flexibility through any one of the first three dimensions and non-technical options such as markets or policy measures can incentivize power generation to match the load [8].

Flexibility options have already been reviewed in the real world [2,8] as well as in models [11,12], but none have yet focused on the local level. At this scope, the spatial and temporal resolution have been highlighted as important factors to consider, particularly with regards to modeling distribution networks and storage, respectively [13,14].

Representing flexibility in models is critical as the unavoidable simplifications made in ESMs lead to an underestimation of flexibility requirements [15]. Technical flexibility in particular has been highlighted as one of the main gaps for models to be better able to answer policy questions [9]. Five main options exist to integrate technical flexibility into models: supply-side options, demand-side management (DSM), storage, networks, and sector coupling. Among these, particularly DSM and storage are frequently implemented, with sector coupling gaining traction [11,12]. Indeed, (residential) DSM and the flexibility provided through sector coupling with electric vehicles (EVs) have the strongest impact amongst different flexibility options in ESMs [16,17]. Nonetheless, explicit supply-side options and distribution networks can also affect results, albeit to a lesser extent [16,18].

Whereas supply-side options inside of the model tend to reduce the degrees of freedom to better approximate the real-world operation of plants, DSM increases them by allowing for either load shifting, whereby the total demand remains the same but the time during which it is met changes, or load shedding, where some share of the demand is not satisfied but curtailed [19]. Important parameters when modeling DSM include how much load can be shifted or curtailed and over what duration [20].

Furthermore, storage is another essential dimension of technical flexibility. Both short-term storage and long-term electricity storage as well as thermal storage are projected to be key components of the decarbonized European energy system [21]. In ESMs, storage technologies are often represented using the simple storage model (SSM), which ignores the physical characteristics of storage technologies [22]. The SSM describes storage modeling using a generic approach that uses different parameters for different storage technologies but otherwise an identical mathematical formulation. Thus, only an energy balance is modeled that is constrained by any combination of a maximum storage capacity, (dis-)charge rate limits, self-discharge, and charging efficiencies.

An important aspect of meaningful storage modeling is temporal resolution [13]. A temporal resolution of typical periods, as frequently used in ESMs, makes it challenging to model long-term storage and could lead to incorrect insights by overestimating the potential of renewables [23–26].

## 2.2 Robustness modeling

Robust insights and decision-support from modeling depend on many aspects, but two main elements are focused on here as these represent two of the most frequently mentioned research gaps in ESMs: the representation of the social dimension and uncertainty assessments [1,27,28].

Krumm et al. [29] reviewed the modeling of the social dimension in ESMs, identifying social acceptance, behavior, actor heterogeneity, public participation and transformation dynamics as the five main aspects that can be included in models. Particularly social acceptance and public participation relate to stakeholder engagement, which is an essential consideration for local ESMs and has frequently been highlighted as one of the most important research gaps on a local level [13,30–32]. Stakeholder engagement also includes the clarity of tools as stakeholders were found to be less likely to use more complex tools [33,34]. Recent research also considers additional options to integrate social acceptance in ESMs such as scenicness ratings [35] or an enforced equitable distribution of renewable generation capacity [36]. In optimization models, it is further important to add heterogeneity in the representation of actors to avoid the so-called "bang-bang effect", whereby a small change in model assumptions results in an outsized impact on the model outputs [37].

A review of uncertainty assessment methods in energy system optimization models by Yue et al. [38] discusses scenario and sensitivity analyses as by far the most frequently used approaches to evaluate uncertainties. These allow to investigate uncertainties regarding the model data and assumptions, but four other approaches are highlighted as they allow to tackle uncertainties more thoroughly: Stochastic Programming, Monte Carlo Analysis, Robust Optimization and Modeling to Generate Alternatives (MGA). Only the last of these allows to confront structural uncertainties, which arise because a model never accurately represents the real-world system [39].

Such approaches are important because minimizing cost, a frequent objective of optimization models, has been demonstrated to be an insufficient proxy to predict real-world energy transitions as decision-makers value other factors beyond cost [40]. Having a broader range of reported metrics and objectives is important to enable informed decision-making that takes into account trade-offs and differing priorities, which is especially important in the context of local planning with a large number of stakeholders [13]. An explicit trade-off assessment is possible through methods such as multi-objective optimization or multi-criteria decision-analysis but reporting a larger number of metrics that might prove relevant to different stakeholders can already facilitate an implicit valuation of different alternatives [41].

### 2.3   Study contribution

A number of reviews have looked at the modeling of flexibility [11,12], social aspects [29] or uncertainty assessments [38] or even both of the latter [9]. However, none of these provide a local scope, a comprehensive overview of all of these dimensions or an identification of best practices.

Meanwhile, reviews of local ESMs are numerous [4,13,14,30–33], but none of these focus on any of the above areas. This is the case even though multiple studies highlight the importance of stakeholder engagement [4,13,32,33] and uncertainty assessments [14,30]. Sector coupling has been suggested for a more efficient implementation of renewables in local ESMs [30,31], echoing Savvidis et al. [9] who find that the representation of technical flexibility in ESMs is lacking relative to the needs for effective decision-support, particularly given a high penetration of intermittent renewables. To the best of the knowledge of the authors, this paper reviews the modeling of both flexibility and robustness in local, integrated ESMs for the first time.

# 3 Methodology

The review process consists of two main steps, the article selection process and the evaluation of the final sample. First, relevant articles were identified using a Scopus[†] search. To this end, different search terms were tested iteratively based on the number and relevance of retrieved articles. The finalized expression includes the local dimension as the scope would have been too extensive otherwise, while anything relating to flexibility or robustness was excluded because several studies that were deemed relevant were not identified in those configurations. It was employed on August 11[th], 2021 to identify 962 results and is as follows:

TITLE ("energ*" AND ("simulat*" OR "model*" OR "optim*" OR "analy*" OR "assess*" OR "system") AND ("region" OR "municip*" OR "communit*" OR "district" OR "cit*" OR "urban*" OR "local" OR "neighb*rhood")) AND TITLE-ABS-KEY ("energ*" AND ("simulat*" OR "model*" OR "optim*" OR "analy*" OR "assess*" OR "system") AND ("region" OR "municip*" OR "communit*" OR "district" OR "cit*" OR "urban*" OR "local" OR "neighb*rhood") AND ("energy system*")) AND ( LIMIT-TO ( DOCTYPE,"ar" ) ) AND ( LIMIT-TO ( LANGUAGE,"English" ) )

To ensure the relevance and comparability of the final sample, five selection criteria were applied as shown in Table 1. The aim of the selection criteria shown in Table 1 is to have a final sample that allows to describe the state-of-the-art modeling of flexibility and robustness in local, integrated ESMs. To this end, the first criterion guarantees that the spatial scope is respected. The next two criteria are used to ensure that the studies are comprehensive in their representation of the energy system and the final two criteria make sure that the studies apply the developed models for insight.

*Table 1: Selection criteria for the article selection process.*

| Selection criterion | Definition |
|---|---|
| **Local** | Spatial scope between a group of buildings and a city |
| **Integrated** | Consideration of at least the thermal and power sectors |
| **Systemic view** | Scoping the entire energy system rather than single technologies/plants or individual actors (e.g. utilities) within a larger context |
| **Investment planning** | System design; studies that only look at dispatch are excluded |
| **Case study** | Real-world application as part of the study |

In a first step, only the abstracts and keywords were screened using the selection criteria, which allowed to reduce the number of articles down to 305. The remaining articles were then analyzed individually and any that did not meet one of the five criteria above during the study evaluation step were further excluded, leading to a final number of 116 articles. The supplementary information (SI) contains a list of articles that were discarded during the second step, flagged with a numerical code classifying the reason for their exclusion.

To systematically assess the selected articles, an attribute catalog was used. Based on Weinand et al. [42], it identifies general information on the model formulation, the implementation of spatial and temporal details and the coverage of different technologies. This catalog was then augmented to capture information on flexibility and robustness modeling based on criteria found in the literature and refined by testing the initial catalog on a random subsample of 15 articles.

The five aspects of technical flexibility introduced in section 2.1 were kept to characterize this dimension in the final attribute catalog, while factors such as markets or regulations were considered outside of the scope of this review due to their diversity. To assess robustness, the criteria from Krumm et al. [29] and Yue et al. [38] discussed in section 2.2 were complemented with criteria on the number and type of objectives and broader metrics which are a pre-requisite for informed decision-making. The concrete criteria that were chosen for these two dimensions are shown in Table 2.

---

[†] https://www.scopus.com/search/form.uri?display=advanced

*Table 2: Criteria in the attribute catalog to assess technical flexibility and robustness. The individual criteria are binary with the exception of informed decision-making, which is quantitative and uncertainty assessment, which is categorical.*

| Technical flexibility | | Robustness | |
|---|---|---|---|
| **Dimension** | **Criteria** | **Dimension** | **Criteria** |
| **Sector coupling** | Combined Heat-and-Power, Heat pumps, Electric Vehicles | **Actor heterogeneity** | Grouping of customers, households or buildings |
| **Storage** | Battery, Pumped hydro storage, Thermal storage, Hydrogen | **Social acceptance** | Trade-off analysis (e.g. MCDA, multi-objective optimization) |
| **Distribution networks** | Heat grid, power grid, gas grid | **Behavior** | Retrofit options, endogenous changes |
| **Demand-side management** | EV modeling, load shifting, load shedding | **Informed decision-making** | Number and type of solutions and reported metrics |
| **Supply-side measures** | Operational constraints, Power-to-X | **Uncertainty assessment** | Deterministic, stochastic, or near-optimal methods |

The results of the final article evaluation can be found in the SI. During the analysis, studies providing details on the implementation of any of the multiple dimensions of flexibility or robustness were flagged. In this way, more detailed information could be collected post-hoc to describe the state-of-the-art modeling of different dimensions of technical flexibility and robustness in local, integrated ESMs. As such, existing solutions to integrate technical flexibility and robustness in models could be described where they are present and research gaps pointed out where solutions are either completely absent or rarely modeled.

## 4   Results

This section presents the results of the analysis. The final selection of 116 studies is first described in section 4.1 to provide context for the consequent sections 4.2 and 4.3, which describe the modeling of flexibility and robustness options respectively. The entire analysis in this section is based on examples from the study sample and the information retrieved using the attribute catalog.

### 4.1   Sample characterization

As Figure 1a shows, more than half of all studies consider not only power demand, but also that for both heating and cooling. When only one thermal sector is considered, this is generally heating demand while just four studies consider cooling without heating, all of them looking at tropical areas [43–46]. The residential sector is most frequently considered as Figure 1b demonstrates. While the commercial sector also has a good representation in the study sample and industry is also included in 40% of all studies despite its heterogeneity, transportation in particular is comparatively lacking. A large majority of studies use optimization models to plan their energy systems as Figure 1c illustrates, most frequently with mixed-integer programming (45%) followed by meta-heuristic approaches (24%), linear programming (15%) and various other methods (16%). A significant minority uses simulation models with just one study not fitting in either box, this one using a life-cycle assessment (LCA) methodology [47].

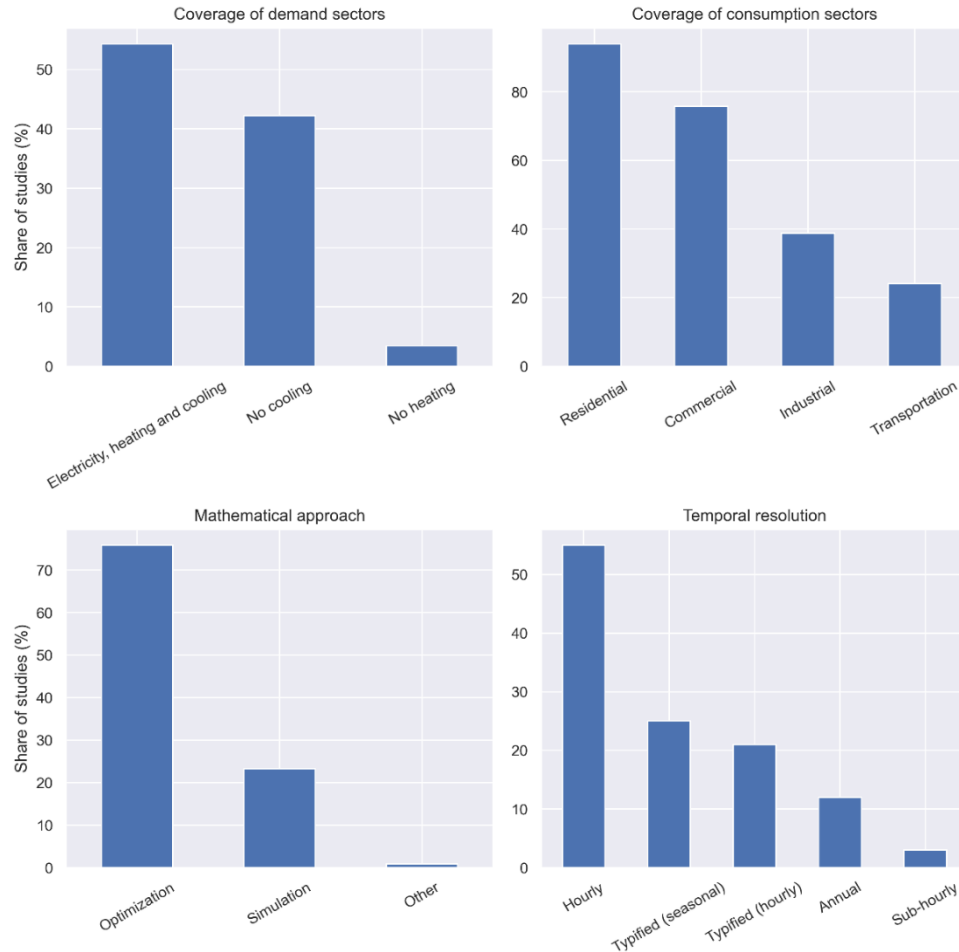

*Figure 1: Overview of the model structure of the reviewed sample of local, integrated ESMs. The panels show the share of studies with respect to (a) the coverage of demand sectors, (b) the coverage of consumption sectors., (c) the modeling approach and (d) the temporal resolution.*

All studies in this sample aim to model at least one year and a quarter even use a multi-year horizon, i.e. model more than one year explicitly. As Figure 1d shows, an hourly resolution is most frequent, followed by a typified seasonal resolution, defined to be anything with less than 72 time slices per year. Such studies can range from just three time slices, usually representing winter, summer and mid-season [48] over multiple time slices per season [49] to 24 time slices to represent each hour of a typical day [45]. The cut-off to a typified hourly resolution was defined to be at 72 time slices as that makes it possible to account for three seasons with an hourly resolution and thus for both seasonal and diurnal variations [50]. This is important to consider variations in the load profile or the generation profile of intermittent renewables such as solar PV or wind.

To analyze the system operation in addition to investment planning, an even higher temporal resolution is favorable, ideally sub-hourly. Such an approach is used by only three studies in the entire sample. While Wills et al. [51] use a simulation-based approach that enables a 5-min resolution and Wilke et al. [52] a genetic algorithm that allows the process of quarter-hourly inputs, another method is employed by Scheller et al. [53], who recursively solve 48h-sized sub-problems with a quarter-hourly resolution to reduce the computational complexity.

Pivoting from the temporal to the spatial scope, the sample remains heterogeneous despite the restriction to a local scope and ranges from energy system planning for a single-digit number of buildings [54,55] to multi-million cities such as Beijing or Shenzhen [56,57]. The spatial resolution can also vary, from single-node municipalities [58] to those considering multiple districts within a municipality [59] and from single-node districts [60] to those where each building is represented individually [61].

The variance in spatial resolution can also provide information on level of technological detail as Figure 2 illustrates. A trade-off is found between the number of technologies and the number of nodes. The latter includes not only the spatial resolution itself but also heterogeneity within the smallest spatial unit, e.g. the number of building types considered per district. Three notable outliers can be found that seem to evade this trade-off. Kuriyan & Shah [62] consider 500 individual nodes in a spatially explicit manner, but only allow for power generation from biomass, simplifying that part of the model to reduce complexity. While McKenna et al. [63] are able to consider four districts with 100 building types each since they only consider 72 time slices per year, Fonseca et al. [64] use a combination of mixed-integer non-linear programming with evolutionary algorithms to overcome the additional complexity from modeling 85 buildings individually.

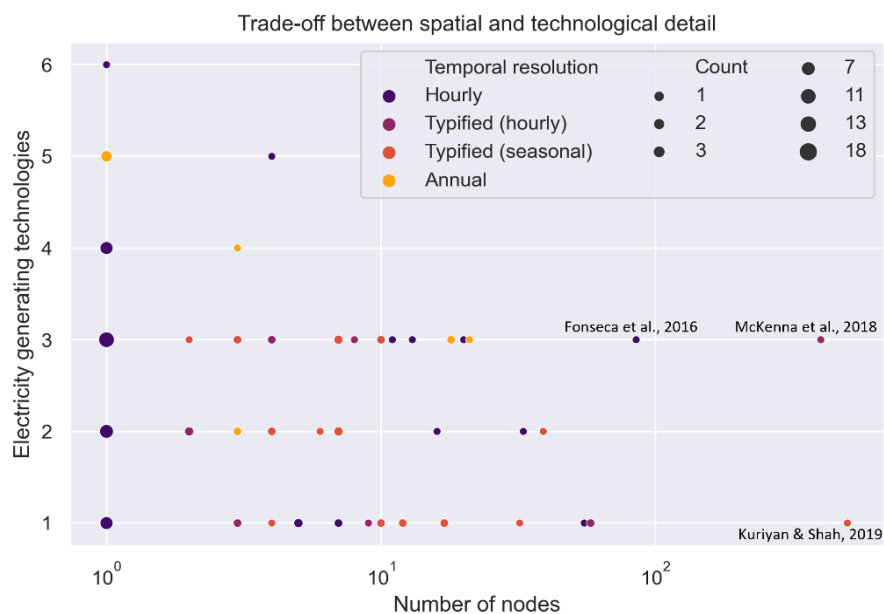

*Figure 2: Trade-off between the number of electricity generating technologies, the number of nodes, and the temporal resolution.*

Notably, the only study considering all six possible power generation technologies uses a simulation model and does not consider all technologies simultaneously [65]. With the exception of Väisänen et al. [47], who use LCA, and Orehounig et al. [66], who use a multi-node approach instead, all of the studies that consider five different electricity generating technologies use a multi-year horizon. All the remaining optimization models also consider large areas as a single node, four times Beijing and once Waterloo [56,67–70] with only Trutnevyte et al. [71] considering a smaller area using a simulation model. Geothermal is the least frequently considered technology for electricity generation and has only been considered in four studies, thrice for cogeneration [72–74]. Instead, it is mainly modeled exclusively as a source of renewable heat [69,70,75–79].

## 4.2 Flexibility modeling

Five different dimensions of flexibility provision in ESMs were analyzed using the flexibility criteria shown in Table 2. All aspects of technical flexibility with the exception of network modeling have been more frequently considered in more recent studes as Figure 3 shows with the most notable increase found for storage modeling. Still, this does not say anything about the level of detail with which these dimensions are represented. Based on the order of their popularity in current modeling practice, this section discusses ways in which the different flexibility dimensions are modeled, concluding with Table 3 highlighting best-practices to model technical flexibility.

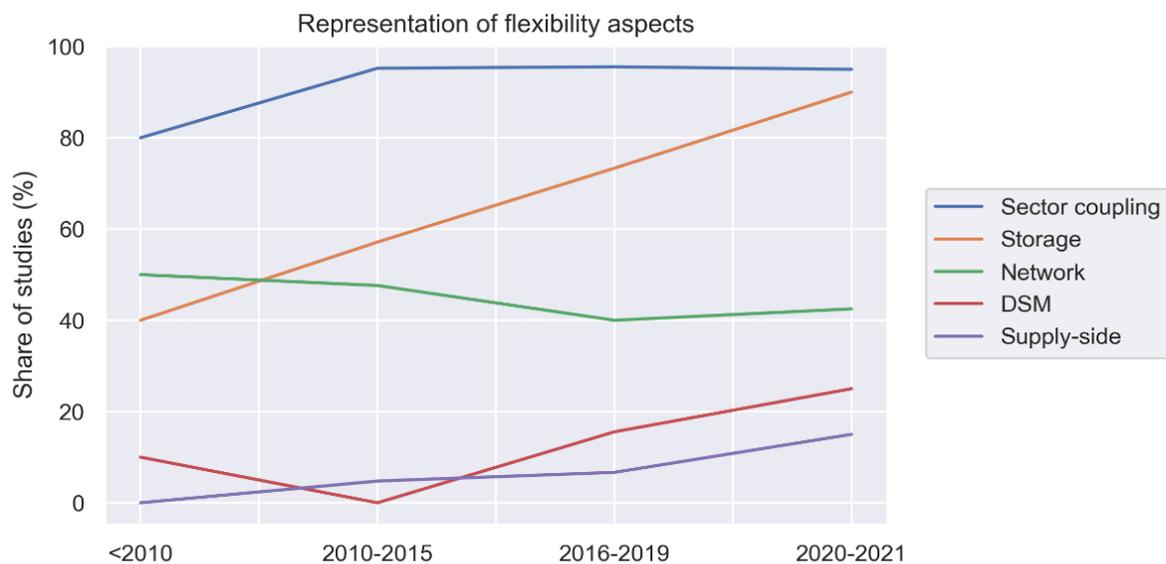

*Figure 3: Frequency with which aspects of technical flexibility have been considered over time in local, integrated ESMs.*

The fact that sector coupling is the most common way in which flexibility potential is tapped in the study sample is unsurprising as it was also a selection criterion. Indeed, more than 90% of studies consider either heat pumps (HP, present in 71% of all studies), Combined Heat and Power (CHP, 67%), or thermal storage (54%), which are the third, fourth and sixth most common technologies among all considered behind solar PV and boilers. There are still a few studies that consider both the thermal and power demand separately, adding complexity to the problem they model but not taking advantage of the potential benefits that could be identified from their combination [80–84].

Storage is the second most frequently modeled aspect of technical flexibility in the sample. Despite this, one-third of all studies do not mention the inclusion of any type of storage technology and even of those that do include it, an additional third do not provide sufficient information to describe how storage modeling is implemented. In a small number of studies, a physical storage model is used, which can represent the physical characteristics of batteries [85], of hydrogen [86], or most frequently of storage tanks used for thermal storage [64,73,87,88].

The remaining half of all studies that do consider storage but not through physical storage modeling can be described using the SSM. As it is technology-agnostic, this approach can be used to model electric (battery) storage, heat or cold storage or even hydrogen [52,66,86,89–91]. Whereas charging efficiencies are included in 80% of all studies that use the SSM and a self-discharge in 60%, rate limits on the (dis-)charge are only included in 30% of these studies and less than 20% enforce balancing over some time horizon.

Also long-term storage can straightforwardly be modeled using the SSM when an hourly model resolution is used [92]. However, models with a typified temporal resolution frequently force daily balancing of storage by adding a constraint to match the storage level at the end of the day to that at its beginning, which does not allow for long-term storage modeling [89,93–96]. Suciu et al. [97] overcome this issue by linking the typical days to real-time slices according to their length.

Distribution networks are the third-most common type of technical flexibility modeled in the sample. Still, the overwhelming majority of all studies that represent heating and power grids use exogenous grids and two-thirds of all studies do not consider any transmission constraints, marking a sharp drop-off in the level of modeling detail compared even to storage. When networks are present, this is often achieved in studies with a small spatial scope by representing each building individually as a node that can export, import, relay or even store energy, potentially using site-specific data on the available links [94,98,99]. Jalil-Vega et al. [100] use a similar approach with districts as nodes instead of buildings whereby the grid length is optimized for intra-node networks and the grid capacity for inter-node networks. Fleischhacker et al. [101] plan power, heating and even gas grids using the tool "rivus" whereby the capacity of each pre-defined link is optimized using assigned costs per $m^2$ and per kW. Further, more complex approaches with algorithms that use, e.g., heating densities exist as well [102–104].

Fourth are options for DSM, which generally work by separating load into one of three categories, either load that is fixed, load that can be shifted or load that can be shed [105]. Price signals are frequently used to constrain the use of DSM in models. Particularly load shedding is typically implemented as the option of last resort with significant penalty costs [92,106,107]. Load shifting in contrast is more flexible. Capone et al. [108,109] allow to shift thermal load by half an hour without assigning any costs and Qiu et al. [110] enable shifting within a day but apply a user satisfaction constraint to limit deviations from the energy demand profile. While Chen et al. [111] assign a cost that depends not only on how much is shifted but also on the duration, Dominković et al. [44] apply a frequency constraint and duration limits to load shifting in addition to a price signal.

Electric vehicles represent a large source of additional energy demand in a decarbonized energy system that can put a strain on it or provide a large degree of additional flexibility if the load is a source of shiftable DSM [87,112]. Nonetheless, only about 20% of studies consider electric vehicles and of these, two-thirds apply a fixed charging profile which means that the potential advantages of vehicle electrification to the grid cannot be considered. Rather than simply charging according to a pre-defined load profile, EVs can be modeled to charge when it is optimal for the system (smart charging) or even as a type of virtual storage (vehicle-to-grid or V2G), both with additional constraints to ensure that sufficient energy is available when the vehicle ought to be used for a trip [102,106,113]. Cao et al. [87] contrast the uncontrolled charging of EVs with smart charging and Heinisch et al. [112] compare both of these options with V2G, both finding benefits from increasing sector coupling, the former economically, the latter in an increased integration of local solar PV in the energy system.

Finally, the least frequently encountered option for technical flexibility in the sample was the modeling of the supply-side, which falls into one of two categories. First, operational constraints can be used to approximate the real-world operation of plants, such as minimum up and downtimes [114], a switching frequency constraint on CHP [94], start and stop costs [115] or ramp limits [87]. Secondly, explicit peaking technologies [116,117] or the conversion of excess renewable generation to other energy carriers (power-to-x) that can more easily be used [104,118–120] allow to add a degree of flexibility to the model. The latter application is most frequent with hydrogen and can be for sector coupling, e.g. to meet mobility demands [121] or even as the foundation for an entire energy system [122], but also just to deal with excess generation, e.g. by providing another input to CHP [123] or using hydrogen as long-term storage [117].

*Table 3: Challenges and best-practice examples related to the modeling of flexibility options at the local level. The share of studies that already use the described approach and references for exemplary implementation are also given.*

| Flexibility challenge | Best practice description | Adoption share | Exemplary implementation |
|---|---|---|---|
| **Sector coupling** | Electric vehicles with V2G | 8% | [87,112] |
| **Storage modeling** | Simple storage model with a combination of charging efficiencies, self-discharge and possibly (dis-)charge rate limits | 40% | [52,66,86,89–91] |
| **Network** | Model network explicitly with constraints | 9% | [94,98–100] |
| **Demand-side management** | Assigning shiftable load with constraints on the quantity and duration | 3% | [108–111] |
| **Supply-side constraints** | Operational constraints such as start-and-stop costs or ramp limits | 9% | [87,94,114,115] |

## 4.3 Robustness modeling

Modeling of robustness is separated into two main areas in this article, the social dimension and uncertainty assessment. The former category includes concepts such as heterogeneity, behavior or social acceptance whereas the latter considers informed decision-making as well as the methods used for uncertainty assessment. These options to integrate non-technical constraints in local ESMs and to strengthen robustness are discussed in this section and recapitulated in Table 4.

To introduce heterogeneity in models and thereby reduce the "bang-bang effect", clustering based on demographic information [96] or energy consumption data [118] have been used in the sample. More frequently however, heterogeneity is based on information with a better availability such as the type of building, its age or size [63,124–126]. If the analysis is sufficiently detailed and small enough in scope, it is also possible to model each building individually [54,84,110,116].

Behavior changes in local ESMs are typically limited to the choice of retrofits, which influence the energy efficiency and thus ultimately energy consumption while no examples were found for endogenous behavior changes. Nearly half of all simulation studies allow for exogenous retrofits, typically through scenarios [46,127–135]. While this is less common, a number of optimization models are also able to account for endogenous retrofits by assigning a cost per m$^2$ or unit of energy that is conserved depending on the data availability [116,118,136–138].

Acceptance is either neglected or addressed exogenously using a multi-criteria decision analysis (MCDA) framework within the studied sample. Using this method, stakeholders evaluate different solutions and rate them based on several metrics to identify the overall most desirable solutions resulting from the modeling process [71]. This method allows for prioritizing between different dimensions such as social, environmental or economic objectives [64]. However, the process has been described as resource-intensive, both for the stakeholders [47] and researchers [63].

Informed decision-making depends on the number and type of objectives and metrics that are used and provided after the modeling process. A first way in which this was assessed is through the number of distinct solutions provided. Here, it was observed that 70% of the sample provide fewer than five distinct solutions, even though optimal solutions are highly dependent on the chosen model assumptions. Looking at the information that is provided for each solution, 80% of all studies report the cost of each solution and 60% the emissions. However, a large gap opens up thereafter as metrics such as imports (23%), the share of renewable generation (14%) or other sources of emissions (7%) are significantly less common. The most comprehensive picture is provided by Arabzadeh et al. [139], who provide 13 resilience indicators, including the ones described above while Samsatli & Samsatli [140] at the other extreme do not provide any information beyond technical parameters on the system configuration.

Not only the reported metrics are of importance in optimization models but especially also the objective that is chosen. As Figure 4 shows, cost minimization is the most frequent objective and considered for more than 90% of studies while even the second most common objective, $CO_2$ emissions, is only present in less than a third of all studies. Other objectives such as a minimization of primary energy consumption [126,128] or non-renewable energy consumption [141] or a maximization of self-generation [142] are only encountered in individual examples.

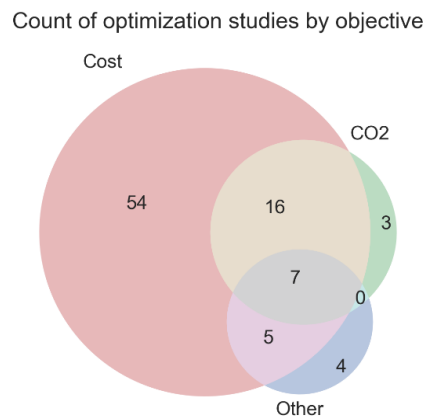

Figure 4: Objectives used in local, integrated energy system optimization models.

Multi-objective optimization is a method to directly assess trade-offs between at least two different objectives, most frequently cost and $CO_2$ emissions as Figure 4 illustrates [143–146]. Notable exceptions also consider aspects such as primary energy consumption [90], self-generation [147,148] or fuel consumption [149,150]. However, some studies also use multi-objective optimization as a substitute to uncertainty assessments, even as these tackle different dimensions of robustness [61,151,152].

The most common type of uncertainty assessment in the study sample is scenario analysis, which was included in over 80% of all studies. Sensitivity analyses were the second most common option, used by a quarter of studies and applied to cost assumptions [48,79,90,143,153], technology availability [117,122,139,154,155], or both [58,93,103,132,156].

While these methods allow to address parametric uncertainties, Modeling to Generate Alternatives (MGA), where the near-optimal solution space is explored, allows to confront structural uncertainties about the limited validity of specific solutions [157]. More sophisticated mathematical formulations such as interval linear programming or chance-constrained programming can also be used to generate more resilient solutions [56,68–70,80,83,106], but require significantly more effort than the simple change of parameters for scenario or sensitivity analyses. Other methods that require less tinkering with the model framework include stochastic sampling of intermittent renewables to account for the uncertainties in their generation profiles [158] or the use of Monte

Carlo analysis to analyze the failure of system components to obtain a resilient, self-sufficientenergy system [150].

*Table 4: Challenges and best-practice examples related to the modeling of robustness options at the local level. The share of studies that already use the described approach and references for exemplary implementation are also given.*

| Robustness challenge | Best practice description | Adoption share | Exemplary implementation |
|---|---|---|---|
| **Actor heterogeneity** | Building typologies, e.g. based on age, size, or use | 27% | [63,125,126] |
| **Social acceptance** | Multi-criteria decision analysis (MCDA) | 3% | [47,63,64,71] |
| **Informed decision-making** | Reporting additional metrics, e.g. imports or share of renewable generation | 32% | [139] |
| **Parametric uncertainty** | Stochastic approaches, e.g. to sample power generation profiles | 7% | [158] |
| **Structural uncertainty** | Modeling to Generate Alternatives (MGA) | 1% | [157] |

# 5   Discussion

The key insights that could be drawn regarding the state-of-the-art modeling of technical flexibility and robustness in local, integrated ESMs as well as the main gaps that were not adequately addressed in the study sample are discussed in section 5.1, followed by a discussion of the review methodology in section 5.2.

## 5.1   Key findings

The average study in this review considers residential and commercial demand for electricity, heating and cooling using an optimization model with an hourly resolution. While it considers sector coupling and storage, it does not consider explicit technology-level flexibility, neither on the supply- nor on the demand-side. These results are similar to the survey-based review of Heider et al. [11], who find the best representation for sector coupling despite the relative recency of the field. However, both supply- and demand-side flexibility are significantly less frequent in our sample, possibly because the description provided within a single study is less exhaustive than what can be collected using a survey.

Even though sector coupling is common, it is primarily found between the electricity and thermal sector while integration of the transportation sector is infrequent. This weakness of local ESMs has previously been highlighted by Keirstead et al. [30] and has become more important since due to the emerging trend of vehicle electrification. Detailed modeling of the vehicle fleet and of V2G can become intricate [159], but methods for complexity reduction [160] or soft-linking of ESMs with other models can help [27]. Network planning studies could similarly be used to generate inputs for local ESMs, which is especially relevant for thermal grids [161]. Here, it is important to note that the share of single-node studies increased from 50% to 70% from pre-2010 to 2020-2021 in Figure 3, which accounts for the decrease in studies which model networks.

A final point relating to technical flexibility is storage modeling, which is completely absent for a third of the sample. Using the SSM represents a low-hanging fruit, but balancing of storage levels has to be handled carefully when long-term storage is modeled in an ESM using a typified temporal resolution [162]. Beyond the choice of the temporal resolution, the technologies that are represented also matter. Kotzur et al. [25] find cost share errors of up to 50% when modeling seasonal storage with typical days, which can be reduced to below 10% when these days are linked. Kannengiesser et al. [163] on the other hand find much smaller effects, a discrepancy that can be explained by the respective shares of solar PV. It is also important to note that both of these studies use at least 120 time slices per year. A significant share of the reviewed studies uses a lower temporal resolution and might thus face more substantial limitations, either because they do not allow for seasonal storage or because they cannot adequately assess it.

Many findings of this study are mostly comparable to those by Krumm et al. [29] as social aspects are mainly accounted for exogenously through clustering prior to modeling, retrofit options and MCDA. A big difference between these two reviews is found in the assessment of heterogeneity. While our results show that it is the best-represented aspect of the social dimension in local ESMs, Krumm et al. found that only agent-based models appropriately accounted for it. This difference can be traced back to different definitions as we only require the presence of different actor groups and they also look for the possibility of interactions between each group. Heterogeneity according to our broader definition is frequently included using building-level information, such as the type, age, or size in our sample. More strictly defined heterogeneity can be modeled through model-linking, but even the broader heterogeneity already represents a big advance as it allows to limit the "bang-bang effect", whereby a small change in model assumptions can have an outsized impact on the modeling results since all actors are assumed to be identical [37].

The way in which uncertainty was assessed in this sample, overwhelmingly using deterministic scenario (80%) or sensitivity (25%) analyses, matches the findings of Yue et al. [38], who found that 75% of all optimization studies used deterministic approaches. More explicit options for trade-off analyses were also explored in MCDA and multi-objective optimization. However, the former is very resource-intensive and the latter forces trade-offs between a limited number of objectives chosen by the modeler. Unmodeled objectives represent a key source of structural uncertainty and while multi-objective optimization allows the modeler to explore the Pareto-optimal frontier, it does not say anything about alternative solutions within the feasible region [164]. MGA is an emerging approach that allows an exploration of the near-optimal space in optimization models systematically to contribute to decision-making beyond providing least-cost solutions and to identify the robustness of solutions [165–168].

## 5.2 Limitations

The methodology was chosen to allow for a reproducible and transparent systematic review that is application-oriented so that the aims of the review can be addressed. A more restrictive search term would have risked missing relevant articles but the chosen expression resulted in the need for filtering to ensure that the studies could be meaningfully assessed on their modeling of technical flexibility and robustness. Since the filtering process was performed manually, no claim is made on the exhaustiveness of the sample.

The attribute catalog was refined multiple times, both before starting the analysis but also during the process. For this reason, but more importantly also because a given paper rarely provides the full model documentation, it is possible that certain aspects, e.g. supply-side constraints, might be more commonly implemented in ESMs than this review identified them to be but just not discussed as they do not contribute to the novelty of the said study. While the general insights regarding the relative frequency of different aspects are not impacted by this, individual studies might be mis- or reclassified.

Some results also have to be interpreted cautiously. The temporal resolution distinguishes between typified seasonal and typified hourly resolutions but did not cover the number of periods per year separately from the number of hours per day, which would have been necessary to identify whether diurnal or seasonal variations are captured in a given study when it uses a typified (seasonal) resolution. Finally, robustness to non-technical constraints is mainly assessed along the social dimension, whereas the integration of market or regulatory constraints into ESMs should be explored in further work.

# 6 Summary and conclusions

The energy system has to undergo a rapid transition to achieve climate neutrality, involving large amounts of decentralized low-carbon technologies to supply energy where it is demanded. Planning this transition requires the consideration of flexibility options to address the intermittency of renewable energy sources, which are expected to contribute significantly to the future energy system. Furthermore, this process also has to tackle non-technical constraints to ensure that the results are robust and valuable for decision-making. A total of 116 studies with real-world applications were analyzed in-depth to assess the representation of technical flexibility and robustness in local, integrated ESMs. This review thereby describes state-of-the-art modeling practice and identifies future research needs.

The coverage of different options to provide technical flexibility has increased within the sample that was analyzed. Sector coupling and storage in particular are considered in virtually all of the most recent studies, while DSM and supply-side options for flexibility are lacking. Still, even though sector coupling is frequent, it is mainly found between the electricity and thermal sector while the integration of transportation, specifically through modeling electric vehicles with V2G, is not considered in models. There are also few constraints on the modeling of storage and networks, which represents another area where improvements in the modeling practice are required.

The robustness of results from ESMs depends on how well they are able to capture socio-political constraints and to assess uncertainty. On the former dimension, heterogeneity can be introduced based on demographic information or more frequently building-level data. The provision of more stakeholder-relevant information and the engagement with that information in systematic processes such as MCDA contribute to stakeholder engagement by promoting acceptance and public participation, which is especially critical on a local level. However, approaches to endogenously consider social factors in models are absent besides theoretical options such as model-linking with agent-based models, which constitutes an important research gap. The same can be said about more systematic uncertainty assessments as these are overwhelmingly performed using scenario or sensitivity analyses. Stochastic approaches or those based on near-optimal solutions (MGA) can be complemented to this end.

Based on the results of this review, we suggest the adoption of the following best practices to model local ESMs. These allow to address technical flexibility and robustness while taking into account the complexity of implementing features and the added value from a higher level of detail in modeling:

1. **Explicitly model electric vehicles**. The shift to electric mobility is a major component of the energy transition and can either massively hamper or increase the flexibility of the energy system, depending on the way in which these vehicles are charged. Models should consider this explicitly using different charging strategies such as V2G or smart charging.
2. **Introduce explicit loads for demand-side management**. In the absence of supply-side flexibility from intermittent renewable energy sources, demand-side options such as load shifting or even load shedding have to contribute. Their modeling should include both temporal and quantitative constraints.
3. **Adopt the SSM when modeling storage systems**. Constraints on the modeling of storage are essential to not underestimate the overall need for flexibility. The SSM allows to apply those in a technology-neutral way well-suited to ESMs using (dis-)charge efficiencies and rate limits as well as a self-discharge.
   4. **Introduce actor heterogeneity in modeling**. The use of homogeneous populations leads to unstable modeling results which are highly dependent on the chosen model assumptions, particularly in optimization models. Local ESMs can take advantage of frequently available building information to deviate from this approach.
   5. **Report a broader range of metrics**. Different decision-makers might not agree on their priorities, which is especially relevant for local planning where a large number of stakeholders are involved. Information beyond cost and emissions on system characteristics such as the share of renewable energy or imports is easily obtainable in ESMs.